\documentclass[usletter, 11pt]{article}

\usepackage{graphicx}
\usepackage{amsmath,amssymb}
\DeclareMathAlphabet\mathbfcal{OMS}{cmsy}{b}{n}% for \mathbfcal to work
\usepackage{amsfonts}
\usepackage{color}
\definecolor{purple}{rgb}{1,0,1}
\definecolor{turquoise}{rgb}{0,1,1}
\usepackage[center, small, sl]{caption}
\usepackage{hyperref}%to include hyper links
\hypersetup{colorlinks,linkcolor=blue,citecolor=blue}%links
\makeatletter
\newcommand{\BIG}{\bBigg@{3}}
\makeatother

\newcommand{\defeq}{\mathrel{\mathop:}=}

\usepackage[section]{placeins}% to be able to use \FloatBarrier

\title{Relativistic Lagrangian and Hamiltonian Description of a Beam with Space-Charge}
\author{Thomas Planche\footnote{tplanche@triumf.ca}, Paul Matthew Jung\footnote{Co-op student, Mathematical Physics, University of Waterloo.} \\~\\ TRIUMF\thanks{This work has been supported by the Natural Sciences and Engineering Research Council of Canada. TRIUMF also receives federal funding via a contribution agreement through the National Research Council of Canada.}}

\begin{document}

    \maketitle
    \tableofcontents

    \section{Introduction}
    We recently became interested in a paper by H.~Qin {\it et al.}~\cite{Qin:2015hta} that describes a symplectic particle-in-cell (PIC) tracking algorithm.
    The algorithm they propose is radically different from usual PIC algorithms used for particle accelerator simulations. In a usual PIC codes, the electromagnetic potential/field is recalculated `from scratch' (from the distribution of macro particles, solving Poisson's equation) at every time step; whereas in their algorithm both the potential/field and the coordinates of macro particles   are pushed self-consistently over each time step using the same symplectic integrator.
    %   H.~Qin {\it et al.} derived from the Hamiltonian of the Vlasov-Maxwell system, with time as independent variable, a set of equations of motion that describe the evolution of a ``discretized'' plasma; discretized in both real space (PIC mesh) and phase space (macro particles). They show how to solve numerically these equations of motion using a first order symplectic integrator.

    In this note we describe the work we have accomplished so far toward implementing a similar algorithm to study space-charge effects in TRIUMF 500~MeV cyclotron.

    \section{Lagrangian}\label{sec:Lagrangian}
    \subsection{Notation}
    In this paper we use notation based on F.E.~Low~\cite{Low282} and  H.~Cendra~{\it et al.}~\cite{Cendra98}.
    All bold symbols are used to denote vectors in $\mathbb{R}^3$. Our beam/plasma is described as a continuous fluid.
    The position in real space of a fluid particle is written:
    \begin{equation}\label{eq:deffX}
        \begin{aligned}
            {\bf x}({\bf x}_1,{\bf v}_1,t)\defeq{\bf x}_1+{\bf\Delta x}({\bf x}_1,{\bf v}_1,t-t_1),
        \end{aligned}
    \end{equation}
    where ${\bf x}_1$ and ${\bf v}_1$ are the position and velocity of the fluid particle at time $t_1$, and ${\bf\Delta x}({\bf x}_1,{\bf v}_1,0)={\bf 0}$.

    $\Phi$ and ${\bf A}$ are the electric scalar and magnetic vector potential. They are related to the electric and magnetic fields through:
    \begin{equation}\label{eq:eb}
        \begin{aligned}
            \mathbfcal{E}=&-\nabla\Phi-\dfrac{\partial {\bf A}}{\partial t};\\
            {\bf B}=&\nabla \times {\bf A}.
        \end{aligned}
    \end{equation}

    $f({\bf x},{\bf v},t): \mathbb{R}^7 \longmapsto \mathbb{R}$ is our plasma/beam density function. The ``initial'' density function is defined as $f_1({\bf x},{\bf v})=f({\bf x},{\bf v},t_1)$ .

    \subsection{Natural Units}
    We choose a system of units such that $\rm{c}=\epsilon_0=\mu_0=1$ (respectively: the speed of light, vacuum permitivity and permeability).
    The conversion of electric charge, vector potential, electric field, mass, time and length to SI units follows respectively:
    \begin{equation}\label{eq:units}
        \begin{aligned}
            q =& \dfrac{q_{_{\rm{SI}}}}{\sqrt{\epsilon_0}} \\
            {\bf A} =& \dfrac{{\bf A}_{_{\rm{SI}}}}{\sqrt{\mu_0}} \\
            \mathbfcal{E} =&  \sqrt{\epsilon_0}\mathbfcal{E}_{_{\rm{SI}}} \\
            m =&  \rm{c}^2 m_{_{\rm{SI}}}  \\
            t =&  \rm{c} t_{_{\rm{SI}}}  \\
            l =& l_{_{\rm{SI}}}. \\
        \end{aligned}
    \end{equation}

    \subsection{Relativistic Space-Charge Lagrangian}
    Based on the Lagrangian proposed by F.E.~Low~\cite{Low282} we write the Lagrangian of a given initial plasma density function $f_1$:
    \begin{equation}
        \begin{aligned}
            &    L  ({\bf x},\dot{\bf x},  \Phi, \dot{\Phi}, {\bf A},   \dot{\bf A};t)  = \\
            &   \iint  f_1({\bf x}_{1},{\bf v}_1)
            \left[-m\sqrt{1-\dot{\bf x}^{2}} + q\dot{\bf x}
            \cdot {\bf A}({\bf x},t)
            - q\Phi({\bf x},t)
            %+ \dfrac{m}{2}|\dot{{\bf x}}- {\bf v({\bf x})}|^{2}
            \right] \,\mathrm{d}{\bf v}_{1}\,\mathrm{d}{\bf x}_1\\
            + \dfrac{1}{2} & \int
            \left[|\nabla\Phi( {\bf x}_{1},t)
            + \dot{\bf A}( {\bf x}_{1},t)|^{2}
            -|\nabla \times {\bf A}( {\bf x}_{1},t)|^{2}\right]\,\mathrm{d}{\bf x}_{1},
        \end{aligned}
    \end{equation}
    where dots stand for partial derivatives w.r.t.~time. The first term is a 6-D integral; it can be seen as a sum over the (continuous) set of fluid particles. The second term is a usual (3-D) volume integral. Note that the projection onto real-space of the 6-D volume of integration coincides with the 3-D volume of integration of the second term: this volume is the volume contained inside the vacuum chamber of our particle accelerator.

    Compared to F.E.~Low's Lagrangian the non-relativistic kinetic energy term has been replaced by $-m\sqrt{1-\dot{\bf x}^{2}}$.

    We will now show how the principle of least action applied to the
    action associated with this Lagrangian:
    \begin{equation}
        \mathcal{S}=\int L \mathrm{d}t,
    \end{equation}
    leads to well-known ``equations of motion''.

    \subsubsection{Variation of the Action in ${\bf x}$}
    Variation of the action (see Appendix~\ref{sec:fd}) in ${\bf x}$ leads:
    \begin{equation}
        \dfrac{\delta \mathcal{S}}{\delta {\bf x}}= f_1
        \left[q\nabla(\dot{{\bf x}}\cdot{\bf A})-q\nabla\Phi -\left.\dfrac{\partial \left(\gamma m\dot{\bf x}+q{\bf A}\right)}{\partial t}\right|_{{\bf x}_1,\,{\bf v}_1}\right],
    \end{equation}
    where $\gamma$ is the Lorentz factor: $\gamma = \dfrac{1}{\sqrt{1-\dot{\bf x}^{2}}}$.\\
    Using chain rule:
    \begin{equation}\label{eq:chainRule}
        \begin{aligned}
            \left.\dfrac{\partial{\bf A}}{\partial t}\right|_{{\bf x}_1,\,{\bf v}_1} &=
            \left.\dfrac{\partial {\bf A}}{\partial t}\right|_{{\bf x},\,\dot{\bf x}}
            + (\dot{\bf x}\cdot\nabla_{\bf x}){\bf A}
            + (\ddot{\bf x}\cdot\nabla_{\dot{\bf x}}){\bf A}\\
            \\
            &=\dfrac{\partial {\bf A}}{\partial t}
            + (\dot{\bf x}\cdot\nabla){\bf A} \,.
        \end{aligned}
    \end{equation}
    The vector algebra identity:
    \begin{equation}
        \dot{{\bf x}}\times(\nabla\times {\bf A})=\nabla(\dot{{\bf x}}\cdot{\bf A})-(\dot{{\bf x}}\cdot\nabla){\bf A},
    \end{equation}
    together with Hamilton's principle of least action:
    \begin{equation}
        \dfrac{\delta \mathcal{S}}{\delta {\bf x}}= 0,
    \end{equation}
    leads to Lorentz' equation:
    \begin{equation}\label{eq:Lorentz}
        m\dfrac{\mathrm{d} \gamma \dot{\bf x}}{\mathrm{d} t}
        = q\left[ -\nabla \Phi - \dfrac{\partial {\bf A}}{\partial t} + \dot{{\bf x}}\times(\nabla\times {\bf A}) \right].
    \end{equation}

    \subsubsection{Variation of the Action in $\Phi$}\label{sec:MG}
    The 6-D integral can be split into two following Fubini's theorem;
    variation of the action in $\Phi$ leads to:
    \begin{equation}
        \dfrac{\delta \mathcal{S}}{\delta \Phi}=
        - \int  q f_1 \mathrm{d}{\bf v}_1 -   \nabla\cdot\left(\nabla\Phi + \dot{\bf A}\right)\,,
    \end{equation}
    where we identify:
    \begin{equation}
        \int  q f_1({\bf x}_1,{\bf v}_1) \mathrm{d}{\bf v}_1=\rho({\bf x}_1,t_1)\,,
    \end{equation}
    to be the local charge density calculated at time=$t_1$.
    Hamilton's principle of least action:
    \begin{equation}
        \dfrac{\delta \mathcal{S}}{\delta \Phi}= 0,
    \end{equation}
    leads (at time=$t_1$) to Maxwell Gauss's equation:
    \begin{equation}
        \nabla\cdot\left(-\nabla\Phi - \dot{\bf A}\right) =  \rho
    \end{equation}\label{eq:MG}

    \subsubsection{Variation of the Action in $\bf A$}
    Similarly the variation of the action in ${\bf A}$ leads:
    \begin{equation}
        \dfrac{\delta \mathcal{S}}{\delta {\bf A}} =\int\limits q\dot{\bf x}f_1\,\mathrm{d}{\bf v}_1-\dfrac{\partial}{\partial t}\left(\nabla\Phi+\dfrac{\partial {\bf A}}{\partial t}\right)-\nabla\times\nabla\times{\bf A}\,
    \end{equation}
    where we recognize:
    \begin{equation}\label{eq:defj}
        \int\limits q\dot{\bf x}f_1({\bf x}_1,{\bf v}_1)\,\mathrm{d}{\bf v}_1={\bf j}({\bf x}_1,t_1)\,
    \end{equation}
    to be the local current density calculated at time=$t_1$.
    Hamilton's principle of least action:
    \begin{equation}
        \dfrac{\delta \mathcal{S}}{\delta \Phi}= 0,
    \end{equation}
    leads (at time=$t_1$) to Maxwell Amp\`{e}re's equation:
    \begin{equation}\label{eq:MA}
        \nabla\times\nabla\times{\bf A}={\bf j}-\dfrac{\partial}{\partial t}\left(\nabla\Phi+\dfrac{\partial {\bf A}}{\partial t}\right)\,.
    \end{equation}

    \section{From Lagrangian to Hamiltonian}\label{sec:Hamiltonian}
    We would like now to apply to our Lagrangian a Legendre transformation to obtain the corresponding Hamiltonian.
    With Lagrangians defined from a Lagrangian densities such as:
    \begin{equation}
        L = \int \mathcal{L}\, \mathrm{d}{\bf x}\,,
    \end{equation}
    the corresponding Hamiltonian density $\mathcal{H}$ can be obtained from:
    \begin{equation} \label{legendre}
        \mathcal{H}(\mathbfcal{P}_i,q_i;t) = \sum_i(\mathbfcal{P}_i \cdot \dot{q}_i) - \mathcal{L}(q_i, \dot{q}, t)\,,
    \end{equation}
    where the conjugate momenta densities are defined by:
    \begin{equation}
        \mathbfcal{P}_i = \dfrac{\partial \mathcal{L}}{\partial \dot{q}}\,.
    \end{equation}
    Here again dots stand for partial derivatives w.r.t.~time.
    The resulting Hamiltonian $H$ then writes:
    \begin{equation}
        H = \int \mathcal{H} \mathrm{d}{\bf x}\,.
    \end{equation}
    In our case the Lagrangian derives from two separate Lagrangian densities:
    \begin{equation*}
        \begin{aligned}
            L   =  \iint \mathcal{L}_{\rm{6D}}({\bf x},\dot{\bf x},  \Phi, {\bf A}) \,\mathrm{d}{\bf v}_{1}\,\mathrm{d}{\bf x}_1 + \int \mathcal{L}_{\rm{3D}}(\Phi, {\bf A},   \dot{\bf A}) \,\mathrm{d}{\bf x}_1.
        \end{aligned}
    \end{equation*}
    Note the variable dependencies of the Lagrangian: dotted terms that appear explicitly in one Lagrangian density do not appear explicitly in the other one. This will allow us to take the Legendre transform of each Lagrangian density independently and combine the resulting Hamiltonians.

    \subsection{Temporal Gauge}
    The lack of explicit dependency on $\dot{\Phi}$ in our Lagrangian implies that $\Phi$ has no canonically conjugated momentum density. To overcome this degeneracy, we choose to fix $\Phi=0$ (as in Ref.~\cite{Qin:2015hta}). This incomplete gauge condition is referred to as temporal gauge (or Weyl gauge).
    Within this gauge the electric field is given by:
    \begin{equation}
        \mathbfcal{E}=-\dfrac{\partial {\bf A}}{\partial t}\, .
    \end{equation}\label{eq:EAtemporal}

    \subsection{First Hamiltonian Density}
    Canonically conjugated pairs associated with $\mathcal{L}_{\rm{6D}}$ are:
    \begin{equation}
        \begin{aligned}\label{eq:momentumX}
            {\bf x}\rm{~and~}&\mathbfcal{P} = \dfrac{\partial \mathcal{L}_{\rm{6D}}}{\partial \dot{\bf x}}=f_1\left(\dfrac{m\dot{\bf x}}{\sqrt{1-\dot{\bf x}^{2}}}+ q{\bf A}\right)\, .
        \end{aligned}
    \end{equation}
    In this case, the Legendre transformation (Eq.~\ref{legendre}) writes:
    \begin{equation}
        \mathcal{H}_{\rm{6D}} = \mathbfcal{P} \cdot {\bf \dot{x}} - \mathcal{L}_{\rm{6D}}.
    \end{equation}
    Substituting leads:
    \begin{equation}\begin{aligned}
        \mathcal{H}_{\rm{6D}} = \dfrac{f_1m}{\sqrt{1-\dot{\bf x}^{2}}}\,.
    \end{aligned}\end{equation}
    To properly transfer into coordinate-momenta space we rearrange for $\bf x$ in terms of $\mathbfcal{P}$ in Equation~\ref{eq:momentumX}:
    \begin{equation}
        \dot{\bf x} = \dfrac{\mathbfcal{P}-f_1q{\bf A}}{\sqrt{f_1^2m^2+(\mathbfcal{P}-f_1q{\bf A})^2}}\,.
    \end{equation}
    After substitution into the Hamiltonian:
    \begin{equation}\begin{aligned}
        \mathcal{H}_{\rm{6D}}({\bf x},\mathbfcal{P};t) = \sqrt{f_1^2m^2+(\mathbfcal{P}-f_1q{\bf A}({\bf x},t))^2}\,.
    \end{aligned}\end{equation}

    \subsection{Second Hamiltonian Density}
    Canonically conjugated pairs associated with $\mathcal{L}_{\rm{3D}}$ are:
    \begin{equation}
        \begin{aligned}
            {\bf A}\rm{~and~}&-\mathbfcal{E}\,.
        \end{aligned}
    \end{equation}
    The Legendre transformation gives:
    \begin{equation}
        \begin{aligned}
            \mathcal{H}_{\rm{3D}} = -\mathbfcal{E}\cdot \dfrac{\partial \bf A}{\partial t} - \mathcal{L}_{\rm{3D}}\,.
        \end{aligned}
    \end{equation}
    Using Eq.~\ref{eq:EAtemporal} and substituting the reduced Lagrangian leads:
    \begin{equation}\begin{aligned}
        \mathcal{H}_{\rm{3D}}({\bf A},\mathbfcal{E};t) = \dfrac{\mathbfcal{E}^2}{2}  + \dfrac{\nabla\times{\bf A}^2 }{2}\,.
    \end{aligned}\end{equation}

    \section{Resulting Hamiltonian}
    Combining the two Hamiltonian densities gives:
    \begin{equation}\label{eq:Hamiltonian}
        \begin{aligned}
            &H({\bf x},\mathbfcal{P},{\bf A},-\mathbfcal{E};t) =
            \iint \sqrt{\tilde{m}^2+(\mathbfcal{P}-\tilde{q}{\bf A}({\bf{x}},t))^2}  \mathrm{d}{\bf v}_1  \mathrm{d}{\bf x}_1 \,
            +  \int  \dfrac{\mathbfcal{E}^2}{2} +\dfrac{\nabla\times{\bf A}^2}{2}\,\mathrm{d}{\bf x}_1\, ,
        \end{aligned}
    \end{equation}
    with $\tilde{m}=f_1m$ and $\tilde{q}=f_1q$.

    Compared to the Hamiltonian used in Ref.~\cite{Qin:2015hta}, this one is truly relativistic, and is fully expressed in terms of canonically conjugated variables; we do not need to use a modified Poisson bracket to obtain equations of motions. This will also allow us to use canonical transformations to re-express the Hamiltonian using Frenet-Serret coordinates.

    \subsection{Equations of Motion from Hamiltonian}
    %\href{https://en.wikipedia.org/wiki/Hamiltonian_field_theory#Equations_of_motion}{Hamilton's equations for continuous systems} 
    Hamilton's equations for continuous systems writes:
    \begin{equation}
        \dfrac{\partial q_i}{\partial t}=+\dfrac{\delta H}{\delta \mathbfcal{P}_i} , \,\,\,\,\,\,\dfrac{\partial \mathbfcal{P}_i}{\partial t}=-\dfrac{\delta H}{\delta q_i};
    \end{equation}
    where $\dfrac{\delta}{\delta}$ stand for functional derivatives(see Appendix~\ref{sec:fd}). From the Hamiltonian given in Eq.~\ref{eq:Hamiltonian} we derive the following equations of motions:
    \begin{equation}\label{eq:eom1}
        \dot{\bf x} = \dfrac{(\mathbfcal{P}-\tilde{q}{\bf A})}{\sqrt{\tilde{m}^2+(\mathbfcal{P}-\tilde{q}{\bf A})^2}}\,,
    \end{equation}
    \begin{equation}\label{eq:eom2}
        \dot{\mathbfcal{ P}} =   \tilde{q}\nabla(\dot{{\bf x}}\cdot{\bf A})\,,
    \end{equation}
    \begin{equation}\label{eq:eom3}
        \dot{\bf A}  =  -\mathbfcal{E} \,,
    \end{equation}
    % %   \begin{equation}\label{eq:eom4}
    % %       - \dot{\mathbfcal{E}}({\bf x}_1,t_1) =    -\nabla\times(\nabla\times{\bf A}({\bf x}_1,t_1))+{\bf j}({\bf x}_1,t_1)\,,
    % %   \end{equation}
    \begin{equation}\label{eq:eom4}
        - \dot{\mathbfcal{E}} =    -\nabla\times\nabla\times{\bf A}+{\bf j}\,,
    \end{equation}
    where dots are partial time derivatives.

    \section{Numerical Integration of the Equations of Motion}

    % \subsection{Spliting the Hamiltonian}
    % Our goal is to integrate equations of motion using a standard symplectic integrator. To do so we will need to get equations of motion from a Hamiltonian written as a sum of analytically solvable parts.
    % This can be achieved assuming, as often with particle accelerator and beam transport lines, that the momentum of beam particle is dominated by one of its component.

    \subsection{External Field}
    To account for the effect of an external magnetic field one has to add an extra vector potential ${\bf A}^{ext}$ to Eqs.~\ref{eq:eom1} and~\ref{eq:eom2}.

    The vertical component of the magnetic field in the mid plane of a cyclotron can in general be written as a sum of Fourier harmonics:
    \begin{equation}
        B_z(r,\theta,z=0)=B_0(r)+\sum_n C_n(r)\cos(n\theta)+S_n(r)\sin(n\theta)\,,
    \end{equation}
    where $r,\theta,z$ are polar coordinates. A corresponding vector potential is given by:
    \begin{equation}
        \begin{aligned}
            A^{ext}_{\theta}(r,\theta,z=0)&=\dfrac{1}{r}\int_0^r rB_z(r,\theta,z=0)\,\mathrm{d}r\,,\\
            A^{ext}_r(r,\theta,z=0)&=A^{ext}_z(r,\theta,z=0)=0\,.
        \end{aligned}
    \end{equation}
    The condition $A_r(z=0)=A_z(z=0)=0$ is referred to as the ``accelerator gauge''~\cite{GordonLecture}.
    The extrapolation of this vector potential to second order in $z$, assuming mid plane symmetry and using Maxwell Amp\`{e}re's equation, leads:
    \begin{equation}
        \begin{aligned}
            A^{ext}_{\theta}(r,\theta,z)&=\dfrac{1}{r}\int_0^r rB_z(r,\theta,z=0)\,\mathrm{d}r+\mathcal{O}(z^4)\,,\\
            A^{ext}_r(r,\theta,z)&=\dfrac{z^2}{2}\BIG[\theta B'_0(r) + r \theta B''_0(r)\\
            +& \sum_n \sin(n\theta)\left(-\dfrac{n C_n(r)}{r}+\dfrac{C'_n(r)}{n}+\dfrac{rC''_n(r)}{n}\right)\\
            -& \sum_n \cos(n\theta)\left(-\dfrac{n S_n(r)}{r}+\dfrac{S'_n(r)}{n}+\dfrac{rS''_n(r)}{n}\right)\BIG]+\mathcal{O}(z^4)\\
            A^{ext}_z(r,\theta,z)&=z r \BIG[  \theta B'_0(r)+\sum_n \dfrac{C'_n(r)}{n}\sin(n\theta)-\dfrac{S'_n(r)}{n}\cos(n\theta)\BIG]+\mathcal{O}(z^3)\,.
        \end{aligned}
    \end{equation}
    In the case of TRIUMF 500~MeV cyclotron those Fourier harmonics are know at a finite number or radii (they have been computed from a cylindrical field survey). To interpolate between those radii we use a standard 1-D spline interpolation, which provide the value of the harmonic as well as its first and second derivatives.

    \subsection{Self Field Interpolation}
\begin{figure}[!ht]
        \includegraphics[width=\linewidth]{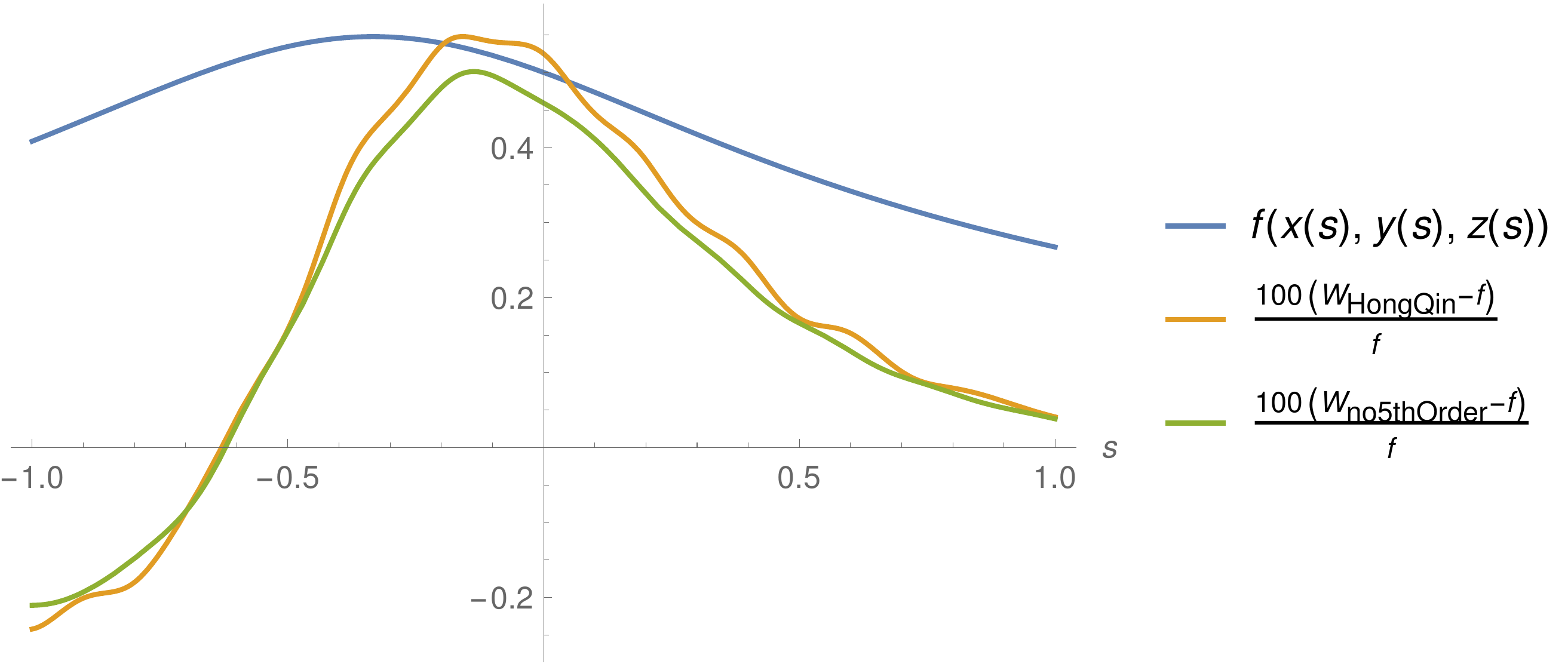}
        \caption{$f(x,y,z)=\frac{1}{\sqrt{x^2+y^2+z^2}}$ is used as a test function to test interpolation. The function is discretized and evaluated along the line  $(x(s)=s+2,y(s)=s,z(s)=2s)$ with two different polynomial kernel interpolators. Relative error of the two interpolators is plotted ($\times 100$) for comparison.
        \label{fig:interp3D3rdOrder}
        }

    \end{figure}

    During the numerical integration of our equations of motion we will keep track of the value of $\bf A$ at only a finite number of location in real-space ({\it i.e.} the nodes of the PIC grid). We will need to interpolate the value of $\bf A$, $\nabla(\dot{{\bf x}}\cdot{\bf A})$, and $\nabla\times\nabla\times{\bf A}$ at arbitrary positions in space. The interpolation of $\nabla\times\nabla\times{\bf A}$ requires an interpolation method which is at least 2$^{\rm{nd}}$-order continuous.
    Hong Qin {\it et al.} proposed to use a 3$^{\rm{rd}}$-order continuous piece-wise polynomial kernel interpolation~\cite{Qin:2015hta}. By studying their kernel function, we found another good candidate:
    \begin{eqnarray}
        W(\mathbf{x}) & = & \mathcal{W}(x/\Delta x)\mathcal{W}(y/\Delta y)\mathcal{W}(z/\Delta z)~,\\
        \mathcal{W}(q) & = & \left\{ \begin{array}{lc}
        0, & q>2~,\\
        -q^8\dfrac{3}{512}+q^7\dfrac{3}{64}-q^6\dfrac{7}{64}+q^4\dfrac{7}{32}-q+1, & 1<q\leq2~,\\ \\
        q^8\dfrac{3}{512}+q^7\dfrac{3}{64}-q^6\dfrac{7}{32}+q^4\dfrac{77}{128}-q^2\dfrac{63}{64}+\dfrac{179}{256}, & 0<q\leq1~,\\ \\
        q^8\dfrac{3}{512}-q^7\dfrac{3}{64}-q^6\dfrac{7}{32}+q^4\dfrac{77}{128}-q^2\dfrac{63}{64}+\dfrac{179}{256}, & -1<q\leq0~,\\ \\
        -q^8\dfrac{3}{512}-q^7\dfrac{3}{64}-q^6\dfrac{7}{64}+q^4\dfrac{7}{32}+q+1, & -2<q\leq-1~,\\ \\
        0, & q<-2~.
        \end{array}\right.
    \end{eqnarray}
    It is also 3$^{\rm{rd}}$-order continuous; it achieves a similar performance on our test function (see Fig.\ref{fig:interp3D3rdOrder}); it is moreover 5$^{\rm{th}}$-order continuous around $q=0$, and slightly more computer-efficient since all 5$^{\rm{th}}$-order coefficients are zero.

    % \section{Change of Independent Variable}
    %
    % The square root term in
    % $\left({\bf  \mathbfcal{P}}-\widetilde{q}{\bf A}\right)^2$ writes differently depending on the choice of coordinate system:
    % \begin{itemize}
    %     \item Cartesian: $(\mathbfcal{P}_x-\widetilde{q}{\bf A}_x)^2+(\mathbfcal{P}_y-\widetilde{q}{\bf A}_y)^2+(\mathbfcal{P}_z-\widetilde{q}{\bf A}_z)^2$,
    %     \item cylindrical: $(\mathbfcal{P}_r-\widetilde{q}{\bf A}_r)^2+(\mathbfcal{P}_{\theta}/r-\widetilde{q}{\bf A}_{\theta})^2+(\mathbfcal{P}_z-\widetilde{q}{\bf A}_z)^2$ (Hagedoorn),
    %     \item Frenet-Serret: $(\mathbfcal{P}_x-\widetilde{q}{\bf A}_x)^2+(\mathbfcal{P}_y-\widetilde{q}{\bf A}_y)^2+\left(\dfrac{\mathbfcal{P}_s-\widetilde{q}{\bf A}_s}{1+x/\rho}\right)^2$ (Courant-Snyder).
    % \end{itemize}

    \FloatBarrier

    %   \subsection{Computation of the Local Current Density}
    %   At every time step the components of the vector $\bf j$ have to be calculated at the nodes of the PIC grid from the distribution of macro-particles. This again is a matter of interpolation. In Eq.~\ref{eq:eom4} $\bf j$ and $\nabla\times\nabla\times{\bf A}$ are added together; since $\bf A$ is interpolated with a 3$^{\rm{rd}}$ order continuous interpolator, $\nabla\times\nabla\times{\bf A}$ is at least 1$^{\rm{st}}$ order continuous. For consistency we choose to compute $\bf j$ using a 1$^{\rm{st}}$ order continuous polynomial kernel interpolation:
    %   \begin{eqnarray}
    %     U(\mathbf{x}) & = & \mathcal{U}(x/\Delta x)\mathcal{U}(y/\Delta y)\mathcal{U}(z/\Delta z)~,\\
    %     \mathcal{U}(q) & = & \left\{ \begin{array}{lc}
    %       0, & q>2~,\\
    %       -\dfrac{q^4}{16}+\dfrac{q^3}{4}-q+1, & 1<q\leq2~,\\ \\
    %       \dfrac{q^4}{16}+\dfrac{q^3}{4}-\dfrac{3 q^2}{4}+\dfrac{5}{8}, & 0<q\leq1~,\\ \\
    %       \dfrac{q^4}{16}-\dfrac{q^3}{4}-\dfrac{3 q^2}{4}+\dfrac{5}{8}, & -1<q\leq0~,\\ \\
    %       -\dfrac{q^4}{16}-\dfrac{q^3}{4}+q+1, & -2<q\leq-1~,\\ \\
    %       0, & q<-2~.
    %     \end{array}\right.
    %   \end{eqnarray}
    %   Result of a test of this lower order interpolation is presented in Fig.~\ref{fig:interp3D1stOrder}.
    %       \begin{figure}[!ht]
    %     \includegraphics[width=\linewidth]{figures/interp3D1stOrder}
    %     \caption{Test of our 1$^{\rm{st}}$ order continuous polynomial kernel interpolation using the same scheme as in Fig.\ref{fig:interp3D3rdOrder}. \label{fig:interp3D1stOrder}
    %   }
    %
    % \end{figure}

    \FloatBarrier

    \section{Appendix}
    \subsection{Functional Derivative}\label{sec:fd}
    Let $\mathcal{C}$ be the configuration space of derivable functions from $\mathbb{R}^n$ to $\mathbb{R}$. Let $\mathcal{S}$ be a mappings from $\mathcal{C}$ to $\mathbb{R}$ (called a functional) such that:
    \begin{equation}
        \mathcal{S}(g)=\int \mathcal{L}(g, \dfrac{\partial g}{\partial x_1}, \dfrac{\partial g}{\partial x_2},.., \dfrac{\partial g}{\partial x_n}) \mathrm{d}x_1\mathrm{d}x_2..\mathrm{d}x_x,
    \end{equation}
    where $g$ is an element of $\mathcal{C}$. The functional derivative of $\mathcal{S}$ with respect to $g$ is defined as:
    \begin{equation}
        \dfrac{\delta \mathcal{S}}{\delta g}=\dfrac{\partial \mathcal{L}}{\partial g}-\dfrac{\partial }{\partial x_1}\dfrac{\partial \mathcal{L}}{\partial \dfrac{\partial g}{\partial x_1}}-\dfrac{\partial }{\partial x_2}\dfrac{\partial \mathcal{L}}{\partial \dfrac{\partial g}{\partial x_2}}..-\dfrac{\partial }{\partial x_n}\dfrac{\partial \mathcal{L}}{\partial \dfrac{\partial g}{\partial x_n}},
    \end{equation}
    see for instance Ref.~\cite{greiner2013field}.

    For the purpose of this paper it is useful to generalize this definition to a configuration space $\mathcal{C}^3$ of functions from  $\mathbb{R}^n$ to $\mathbb{R}^3$; in this case $\mathcal{S}$ becomes a mappings from $\mathcal{C}^3$ to $\mathbb{R}$. Let now $g \in \mathcal{C}^3$ such as:
    \begin{equation}
        g(x_1,x_2,..,x_n)=\begin{pmatrix}g_1(x_1,x_2,..,x_n)\\ g_2(x_1,x_2,..,x_n) \\g_3(x_1,x_2,..,x_n)\end{pmatrix},
    \end{equation}
    then:
    \begin{equation}
        \dfrac{\delta \mathcal{S}}{\delta g}=\begin{pmatrix}\dfrac{\delta \mathcal{S}}{\delta g_1}\\ \dfrac{\delta \mathcal{S}}{\delta g_2}\\ \dfrac{\delta \mathcal{S}}{\delta g_3}\end{pmatrix}.
    \end{equation}

    %\bibliographystyle{/home/tplanche/utils/latex/formating/elsarticle-num}
    %\bibliography{/home/tplanche/utils/latex/Mybib}

\end{document}